\documentclass[twocolumn]{aastex631}

\newcommand\nbluegreen{359}
\newcommand\nviewerdetected{359}

\newcommand\norigdetected{298}

\newcommand\nblue{94} 

\newcommand\ngreen{265}
\newcommand\nvieweroriginaldetected{265}

\newcommand\nred{33}
\newcommand\noriginaldetectedonly{33}

\newcommand\gridbeams{41342}
\newcommand\otherbeams{1774}
\newcommand\totalbeams{43116}
\newcommand\gridpointings{34869}
\newcommand\otherpointings{440}

\newcommand\newlmcpulsar{J0540$-$69}

\newcommand\lmcpulsarperiod{0.909~s}

\newcommand\lmcpulsardm{80~pc cm$^{-3}$}

\definecolor{red}{rgb}{0.47, 0.05, 0.05}
\newcommand{\hl}[1]{{{#1}}}

\begin{document}

\title{Reprocessing of the Parkes 70-cm Survey and Discovery of a New Radio Pulsar in the Large Magellanic Cloud}

\correspondingauthor{Wenke Xia}
\email{wenke.xia@mail.mcgill.ca}

\author[0009-0009-9343-4193]{Wenke Xia}
\affiliation{Department of Physics and Astronomy, Franklin and Marshall College, P.O. Box 3003, Lancaster, PA 17604, USA}
\affiliation{Department of Physics, McGill University, 3600 rue University, Montreal, QC H3A 2T8, Canada}
\affiliation{Trottier Space Institute, McGill University, 3550 rue University, Montreal, QC H3A 2A7, Canada}

\author[0000-0002-2578-0360]{Fronefield Crawford}
\affiliation{Department of Physics and Astronomy, Franklin and Marshall College, P.O. Box 3003, Lancaster, PA 17604, USA}

\author[0000-0002-7700-3379]{Shinnosuke Hisano}
\affiliation{Kumamoto University, Graduate School of Science and Technology, Kumamoto, 860-8555, Japan}

\author{Tai Jespersen}
\affiliation{Department of Physics and Astronomy, Franklin and Marshall College, P.O. Box 3003, Lancaster, PA 17604, USA}

\author{Melanie Ficarra}
\affiliation{Department of Physics and Astronomy, Franklin and Marshall College, P.O. Box 3003, Lancaster, PA 17604, USA}

\author{Mckenzie Golden}
\affiliation{Department of Physics and Astronomy, Franklin and Marshall College, P.O. Box 3003, Lancaster, PA 17604, USA}

\author{Mia Gironda}
\affiliation{Department of Physics and Astronomy, Franklin and Marshall College, P.O. Box 3003, Lancaster, PA 17604, USA}

\begin{abstract}
We have reprocessed the data archived from the Parkes 70-cm pulsar (PKS70) survey with an expanded DM search range and an acceleration search. Our goal was to detect pulsars that might have been missed in the original survey processing. Of the original 43842 pointings, \gridpointings\ pointings were archived, along with \otherpointings\ additional pointings for confirmation or timing. We processed all of these archived data and detected \nbluegreen\ known pulsars: \ngreen\ of these were detected in the original survey, while an additional \nblue\ currently known pulsars were detected in our reprocessing. A few among those \nblue\ pulsars are highly accelerated binary pulsars. \hl{Furthermore, we detected \hl{5} more pulsars with DMs higher than the original survey thresholds, as well as 6 more pulsars below the nominal survey sensitivity threshold (from the original survey beams with longer integrations).} We missed detection of \nred\ (of the \norigdetected) pulsars detected in the original survey, in part because portions of the survey data were missing in the archive \hl{and our early stage candidate sifting method}. We discovered one new pulsar in the re-analysis, PSR \newlmcpulsar\, which has a spin period of \lmcpulsarperiod\ and resides in the Large Magellanic Cloud (LMC). This new pulsar appeared in three PKS70 beams and one additional L-band observation that targeted the LMC pulsar PSR B0540$-$69. The numerous pulsar detections found in our re-analysis and the discovery of a new pulsar in the LMC highlight the value of conducting multiple searches through pulsar datasets.
\end{abstract}

\keywords{Surveys (1671), Radio pulsars (1353)}

\section{Introduction and Motivation} \label{sec:intro}

\subsection{PKS70 Survey and Data Archive}

The Parkes 70-cm (PKS70) Pulsar Survey was conducted at the 64-m Parkes telescope (``Murriyang'') in the 1990s, covering the southern sky visible from the telescope. The survey observations were centered at 436~MHz with a bandwidth of 32~MHz split into 256~channels. The data were sampled at 0.3~ms, making the survey not very sensitive to millisecond pulsars (MSPs). Typical integrations were 157~s per pointing, and the limiting flux density in the original survey was about 3~mJy (but falls off significantly for MSPs and at low galactic latitudes; see Fig. 6 of \citealt{mld+96}). These observing parameters are listed in Table \ref{tbl-1b}. 

In the original survey publications \citep{mld+96, lml+98, dsb+98}, the authors reported observing and searching 43842 tiled, blind survey beams out of the total originally planned gridded survey area of 44299 beams that were to cover the southern sky. A total of \norigdetected\ pulsars were detected, including 101 new pulsars and 197 pulsars that were known at the time. 

Of these \hl{gridded} survey observations, 41342 were archived, pointing at 34869 different positions. The remainder of the data was likely lost or was stored on damaged tapes, but its fate is uncertain (D. Lorimer, private communication). \hl{A small fraction (2\%) of the gridded observations had an integration time that was noticeably longer than the nominal integration of 157~s. The cause could be that the original survey observed some positions with different integrations, or it could be because these were makeup observations of corrupted or bad observations.} In addition to these gridded survey beams, 1774 observations with 440 different targeted positions (some of which were the same as the gridded positions, most likely candidate confirmation beams or timing observations) were recorded as part of the survey and were archived in the project. These additional targeted beams are mostly recorded at 436~MHz (86\% of the targeted beams, with 0.125~MHz channel width) and 1400~MHz (14\% of the targeted beams, with 2-5~MHz channel width). Their integration time ranged from a few seconds (likely test observations or corrupted data) to $\sim$12~hours, targeting the Small Magellanic Cloud (SMC) and Large Magellanic Cloud (LMC); despite the large range, most observations were 5-10 minutes. With the original survey and additional observations, the complete archive contained 43116 observations (35279 pointing positions, representing 80\% of the original pointing positions) available for reprocessing.

\begin{figure*}
    \centering
    \includegraphics[width=\linewidth]{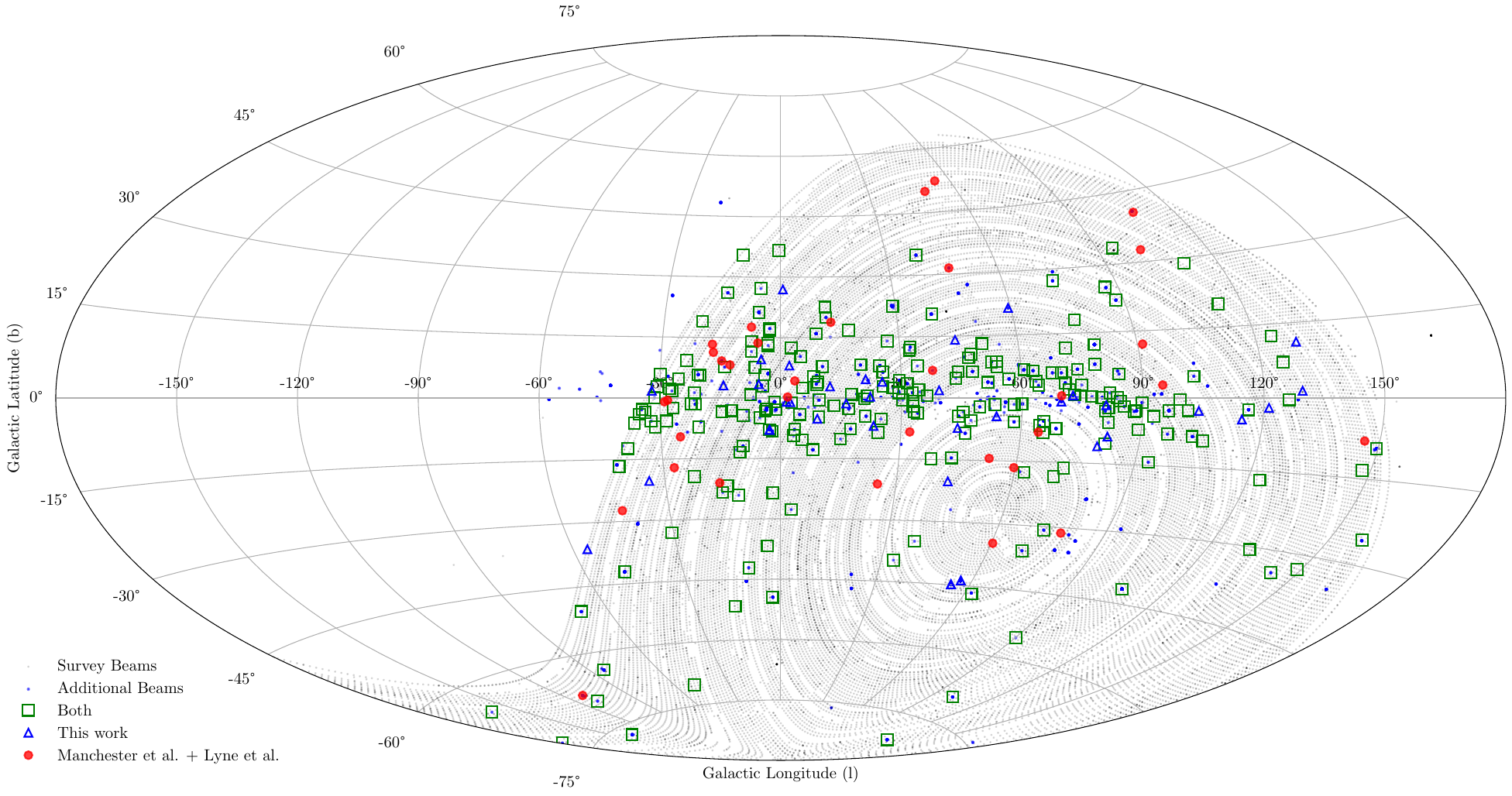}
    \caption{Sky plot of the PKS70 survey beam coverage from archival survey beams (small gray circles) and additional observations (small blue circles), shown in Galactic coordinates. The \hl{white patches} in the sky coverage indicate where the survey beam data are missing in the archive. Pulsars detected by both the original survey and our reprocessing are shown as green squares, while pulsars detected only in our reprocessing are shown as blue triangles. Pulsars detected in the original processing but not in our reprocessing are plotted as red circles (see also Tables \ref{tbl-1} and \ref{tbl-1a}.)}
    \label{fig:skymap}
\end{figure*}

Fig. \ref{fig:skymap} shows the sky coverage of the full archival survey beams in gray. 
Note that this does not include the missing 20\% of the survey that was not archived, which mainly corresponds to the white patches within the survey region. The additional beams are also shown in blue dots; some of these lie outside of the nominal survey region, while many lie within that region \hl{without following} the pattern of the survey beam map.

\subsection{New Processing of the Survey Archive}

We downloaded raw data from the CSIRO Data Archive\footnote{\url{https://data.csiro.au}} and reprocessed the entire survey archive to search for previously missed pulsars and fast radio bursts (FRBs). A search for single pulses and FRBs in the survey data had not been conducted previously. We completed this search prior to the periodicity search described here. The search resulted in the discovery of four new, long-duration ($> 50$ ms) FRBs, which were reported by \citet{chg+22}.

The reprocessing described here for new periodic pulsar signals is motivated by two factors. The first is that the newer software we are using for the periodicity search \cite[PRESTO;][]{r01,rem02} was not available at the time of the original survey. This has better radio frequency interference \hl{(RFI)} excision and visualization features than the older software used in the original data processing. Thus, there may be pulsars in the data that were missed. The second is that, with the vastly increased computational power available now compared to what was available at the time of the original survey, we can explore an expanded search parameter space within a reasonable amount of computational time. We searched up to a much larger maximum DM (3000~pc~cm$^{-3}$, compared with the maximum DM of 777~pc cm$^{-3}$ in the original processing), which could reveal unexpectedly high-DM pulsars in the data. \hl{We also used PRESTO's built-in dedispersion, a modern dedispersion algorithm in contrast to the algorithm from the origin survey that used a linear approximation in the first stage of dedispersion \citep{mld+96, lml+98, taylor74}. The newer dedispersion algorithm is more computationally intensive but results in improved sensitivity at higher DMs, as it avoids the additional smearing effect caused by the approximation.} \hl{Lastly, the} increased computational power allowed us to conduct a search for accelerated (binary) pulsars, which was not done in the original processing. This allowed us to maintain sensitivity to even highly accelerated binary systems, which could have been missed in the prior analysis. Such systems are rare but important to study, as they are ideal systems for testing theories of gravity and gravitational waves \citep{taw+89,wet+03,wet+05}. 

In this paper, we report on our analysis of the survey data, discuss both the pulsar re-detections and those that were missed, and report on the discovery of a new pulsar in LMC, PSR \newlmcpulsar, which appeared in 3 targeted observations of the X-ray pulsar PSR B0540$-$69 \citep{shh+84} in the PKS70 survey archive and 1 other non-survey L-band LMC observation.

\section{Search Methods and Analysis}

We processed all the archived data available from the PK70 archive, including both gridded and targeted beams. 
We first searched each beam for RFI and masked it prior to analysis using the rfifind script in PRESTO. A time window of 1~s was specified while the default values were used for the rest of the parameters. The mean (and median) fraction of data flagged and masked as RFI for the analyzed files was 10\%, but this ranged widely across the dataset (from 0\% to 99.96\%). \hl{ Only 205 (about half a percent) of the collection of beams had RFI flagging of more than 50\%. Of these 205, all but 29 were re-observed successfully.} 

\begin{figure}
    \centering
    \includegraphics[width=1.0\linewidth]{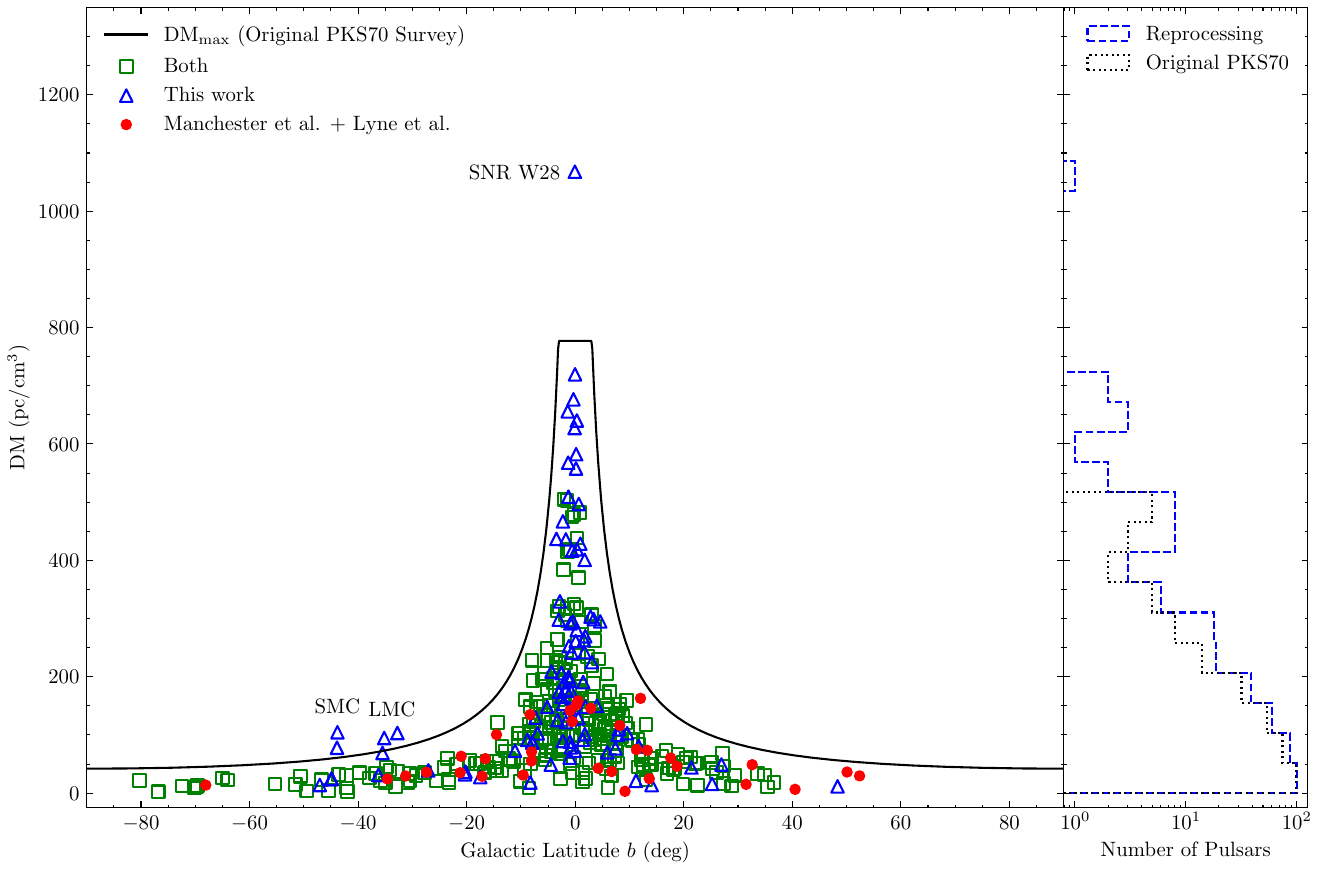}
    \caption{DMs of detected pulsars in both the original survey and this new reprocessing. The left panel shows the DM vs. galactic latitude $b$ of known pulsars detected in the original survey (red circles and green squares) and in this new reprocessing (blue triangles and green squares), along with the maximum DM searched in the original survey (black line). With a higher \hl{maximum DM} (up to 3000~pc/cm$^3$), 5 additional known pulsars with excessive DMs were detected in this reprocessing. The excess DM of all 5 pulsars is attributed to their local environments -- 4 are located in the SMC or LMC, and 1 is associated with a known supernova remnant, as noted in the figure. The right panel shows the distribution of the DMs of all the pulsars detected in the original PKS 70 survey (black) and in this reprocessing (blue). The DM distribution of pulsars detected in the reprocessing expands into higher DM values \hl{compared to the original survey}. }
    \label{fig:dm-galactic-b}
\end{figure}

We then dedispersed each beam into 1453 separate DM steps between 0 and 3000~{pc cm$^{-3}$} in several stages. The DM spacing, downsampling, and range of DMs used in each stage were obtained by using the DDPlan.py script in PRESTO (see Table \ref{tbl-ddplan}). This DM range is significantly larger than the prior search and goes beyond the Galactic electron layer in all directions \citep{cl02}. In contrast, to reduce processing time, the previous search dedispersed the data with a direction-dependent maximum DM, which was the smaller of 777~pc~cm$^{-3}$ and $42/\sin b$~pc~cm$^{-3}$, where $b$ is the Galactic latitude \citep{mld+96}. The greater DM range in this reprocessing allowed for the detection of pulsars with DMs larger than the expected maximum DM value in case of local regions of plasma between us and the pulsar (see Fig. \ref{fig:dm-galactic-b}).

\begin{figure*}
    \centering
    \includegraphics[width=\linewidth]{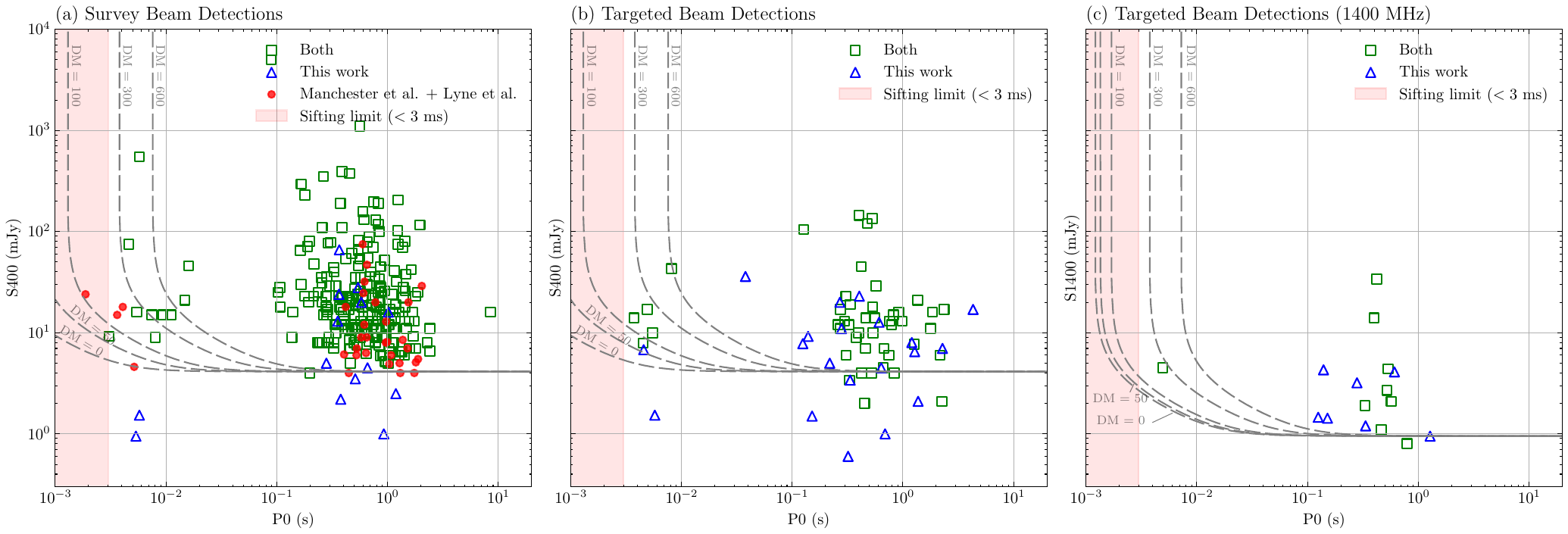}
    \caption{Flux density vs. spin period of known pulsars in the region observed; pulsars detected from at least one survey beam are shown in panel (a), and the pulsars detected only from targeted beams are shown in panel (b) and (c).  In panel (a), sensitivity curves are shown for representative DMs (0, 50, 100, 300, and 500~pc~cm$^{-3}$) and are assuming a duty cycle of 0.08, a detection S/N of 8, and a digitization loss constant $\beta$ of 1.5 in the radiometer equation \citep{mld+96, lml+98}. They also include the effects of dispersive smearing within channels and the sampling time (but not the effects of interstellar scattering), which degrade the sensitivity at small periods. Several pulsars (blue triangles) were detected in our re-analysis that have flux densities below the nominal survey limit, including two with spin periods of less than 10~ms. \hl{These detections were made from observations with longer integrations} In panel (b), the same curves are plotted along with known pulsars that were detected only in an additional targeted beam. Some pulsars are only detected in a targeted beam centered at higher frequencies (L-band) and are plotted in panel (c). We also estimated the sensitivity curves for L-band detections using a sampling time of 1.2~ms, a \hl{central} frequency of 1400~MHz with 5~MHz resolution, and a typical integration time of 300~s. The remaining assumptions are the same as those in the calculation at 400 MHz. 
    Our candidate sifting process removed candidates with periods under 3 ms, and this range is shown by the shaded region. 
    } 
    \label{fig:sens-period}
\end{figure*}

\begin{figure}
    \centering
    \includegraphics[width=\linewidth]{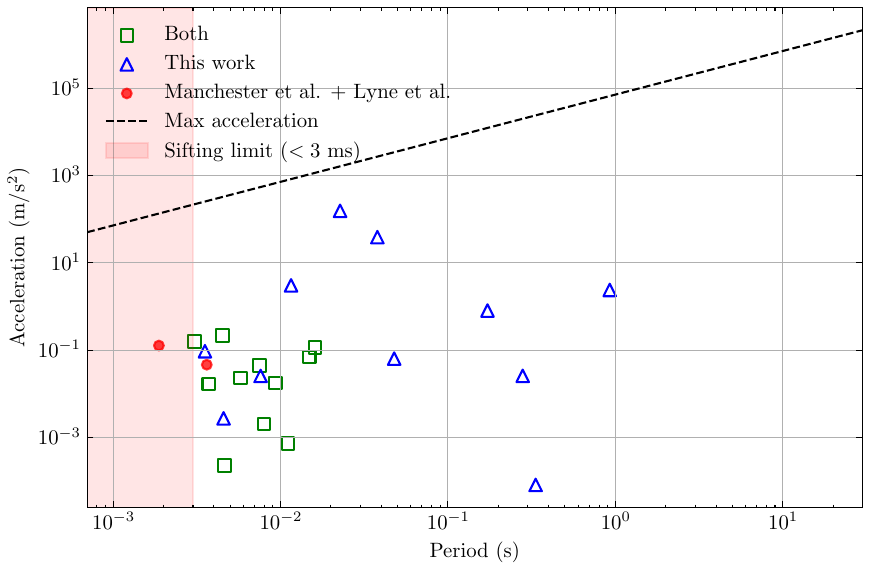}
    \caption{The maximum expected acceleration during the course of the binary orbit is plotted for the set of binary pulsars that were detected. The dashed line indicates the acceleration search limit in the reprocessing for which full sensitivity to accelerated signals was maintained. This limit increases linearly with spin period. The shaded region at periods below $P = 3$~ms indicates where our processing was not sensitive owing to our candidate sifting and removal. The many isolated pulsars detected in the survey (with no measured acceleration in the ATNF pulsar catalog, \citealt{mht+05}) are not plotted. The original survey processing did not employ an acceleration search, and the 13 accelerated pulsars detected in that search are plotted as green triangles (11) and red circles (2). A significant number (11) of additional accelerated pulsars were detected in our reprocessing but were not found in the original survey (blue triangles). We detected almost all of the original detections (11 out of 13) plus a comparable number (11) of new detections not detected in the original search, indicating the importance of this accelerated search. The two pulsars originally detected but missed by us (PSR J0034$-$0534 and PSR J1911$-$1114; red circles) have millisecond periods (and one is below our 3~ms sifting limit), which is likely the reason we missed them \hl{(see Section \ref{sec:missed-pulsars})}.}
    \label{fig:accel-period}
\end{figure}

Each dedispersed time series was high-pass filtered and searched for periodicities with an acceleration range that corrected for a Fourier power spectrum bin drift of 50 bins over the course of the integration. The maximum acceleration to which this corresponds is $a_{\rm max} = z_{max} c P / h T^{2}$, where $z_{max}$ is 50, the maximum bin drift, $h$ is 8, the number of harmonics used in the search, $T$ is 157 s, the typical integration time per beam (see Table \ref{tbl-1b}), $c$ is the speed of light, and $P$ is the pulsar's spin period. We used the native 0.3~ms sampling in the search, which maintained good sensitivity to pulsars with spin periods greater than about 3~ms for DMs $\la 100$~pc~cm$^{-3}$ (see Fig. \ref{fig:sens-period}). Our search maintained full sensitivity to pulsars with a maximum acceleration of 230 m s$^{-2}$ (for a 3~ms spin period), making this search sensitive to highly accelerated binaries (see Fig. \ref{fig:accel-period}). Note that the pulsar with the highest currently known acceleration, the double neutron star PSR  J1757$-$1854 \citep{cck+18}, with a spin period of 21.5 ms, has a maximum periastron acceleration of 684 m s$^{-2}$ and is within the range where it would have been detectable in our acceleration search (although its flux density is too low for detection in this survey by at least an order of magnitude). The periodicity candidates identified from the search were then collected and sifted. The sifting process involved removing redundant candidates in each beam, based on their periods and DMs. Only the candidate with the strongest S/N in the redundancies was retained. In addition, any candidates appearing in fewer than 10 DM trials or with a spin period less than 3~ms were removed in order to reduce the total number of final candidates to investigate further. Note that this period restriction prevented us from detecting pulsars having periods less than 3~ms in the blind search. The period and DM for each candidate were used to refold the data at periods and DMs near these nominal values using the prepfold package in PRESTO. 

Our reprocessing resulted in a total of 364087 candidates. To inspect such a large number of candidates, we developed a web-based plot visualization and rating tool for displaying candidate plots and cataloging classifications from these plots. Teams of undergraduate students from several institutions that are part of the NANOGrav Physics Frontiers Center\footnote{\url{https://nanograv.org}} were engaged in classifying the candidates. A total of 64 students from 9 institutions rated these plots. To identify known pulsars, we used Pulsar Scraper\footnote{\url{https://pulsar.cgca-hub.org/search}} \citep{kap+22} to search for coincident periods (or harmonics of periods), DMs, and coordinates within the Parkes beam size. Candidates with corresponding entries were classified as detections of known pulsars.

\vspace{1cm}
\section{Results and Discussion} \label{sec:res}

We blindly detected \nviewerdetected\ known pulsars in our processing, where 256 pulsars were from gridded beams and 103 pulsars were from targeted beams. \hl{The original survey indicated that all \norigdetected\  pulsars were found in the gridded beams \citep{mld+96, lml+98}, but our comparison of pulsar positions with the gridded and targeted beams shows coincidences that suggest some were detected in the targeted beams. Furthermore, we did not detect 118 pulsars in the survey beams that were reported in the original survey. These pulsars represent $\sim$40\% of the original detections and were detected in the remaining targeted beams instead. Since our reprocessing should be at least as sensitive as the original survey, we speculate that these targeted beams were actually part of the original survey coverage.} Thus, we compare our results from the reprocessing of both the gridded survey beams and the extra beams with the original results.

\subsection{Additional Pulsars Detected in the Survey Re-processing}

Of these \nviewerdetected\ pulsars we detected, the original survey did not detect \nblue\ of them. One reason for some of these additional detections is the use of our acceleration search up to $z_{\rm max} = 50$. Fig. \ref{fig:accel-period} shows the maximum expected acceleration during the course of an orbit (derived from the orbital parameters) as a function of spin period for highly accelerated pulsars (binaries with low maximum accelerations, below $10^{-4}$ m/s$^{2}$, are not included). 

To gauge the detectability of accelerated pulsars in a non-accelerated search, we did a quick test with the 22 binary pulsars shown in Fig. \ref{fig:accel-period}. We compared the strength of the signal in the power spectrum, where zmax was 0 and 50. All of the pulsars appeared as candidates in both of these cases when the data were dedispersed at the known pulsar’s DM. However, three pulsars had significant signal reductions in the spectrum from zmax of 0 compared to the one from zmax of 50 (the sigma value was reduced to 46\% for PSR B0021$-$72E, 84\% for PSR J0737$-$3039A, and 23\% for PSR J2051$-$0827). In other words, these three pulsars exhibited significant acceleration during the observation, whereas the others did not. It is worth noting that PSR J2015$-$0827 was detected in the original survey, despite being highly accelerated. The plotted points in Fig. \ref{fig:accel-period} represented maximum possible accelerations (estimated from the orbital parameters), so it is not surprising that the accelerations actually observed were smaller. This suggests that, although the majority of pulsars did not exhibit increased detectability from the acceleration search, at least some would have had better detectability; thus, it was an important parameter space to search, given the availability of computational resources.

We also detected five high DM pulsars that exceed the maximum DM searched in the original survey. The excess DM for all five pulsars is due to their local environments: 4 in the SMC or LMC, and 1 in a known supernova remnant, as shown in the left panel of Fig. \ref{fig:dm-galactic-b}. Furthermore, we detected several high DM pulsars near the galactic center that the original survey did not detect; the original survey only detected pulsars up to $\sim 550$~pc~cm$^{-3}$ despite their maximum DM limit of $\sim 777$~pc~cm$^{-3}$, whereas we detected several beyond what the original survey detected (see the right panel of Fig. \ref{fig:dm-galactic-b}). \hl{The reason for this difference at these large DMs is still uncertain. We found that it is not related to the DM step size used, pulse profile morphology, or observation integrations. One possible reason is that  a linear dedispersion approximation was used in the initial sub-banding stage in the original processing, which could have resulted in larger DM smearing effects. However, answering this question fully would require more testing and comparison of the approximated dedispersion with the PRESTO dedispersion, which could be addressed in future work.}

Finally, the population of pulsars that we detected extends to lower luminosities than those detected in the original survey. \hl{We detected 11 pulsars below the nominal survey sensitivity threshold (see Fig. \ref{fig:sens-period}), but these detections are all from gridded beams with longer integrations.}

\subsection{Missed Pulsars in the Survey Reprocessing} \label{sec:missed-pulsars}

Despite searching targeted beams in addition to survey beams, we missed \nred\ pulsars that were detected in the original survey. These are listed in Table \ref{tbl-2} along with their refolding status using cataloged parameters and the proposed reason for non-detection (also discussed below).

Of these \nred\ pulsars, we found that 14 of them have no nearby archival beams within 1 beam size. By cross-checking the position of these 14 pulsars, the available beam positions, and the original survey pointing map \citep{mld+96}\footnote{ We reconstructed the pointing map using the procedure described by \cite{mld+96}. }, we found that all of them are possibly located next to a missing beam (listed as ``missing beams" in the table).

We folded the remaining 19 pulsars with PRESTO \citep{r11} using their nearby observations within 1 beam size and parameters listed in the ATNF pulsar catalog \citep{mht+05}\footnote{We used PSRCAT version 2.5.1 for the catalog data, and the same version was used consistently throughout the rest of the paper.}, and we didn't detect 7 of them (listed as ``no detection after folding" in the table). This could be due to (a) a different RFI mitigation method/configuration in the processing, or (b) the original survey revisited these pointings and detected these pulsars, but the revisited observation was not archived.

Finally, there are 12 pulsars whose signals are present in our data but were not blindly detected (highlighted in bold in Table \ref{tbl-missed-pulsar-parameters}). Among them, four pulsars had low detection S/N as reported in the original surveys \citep{mld+96, lml+98}, and two have cataloged 400~MHz flux densities \citep{mht+05} close to our detection threshold. Additionally, 7 of these pulsars are located near the edge of the beam (this includes the pulsars whose closest beam is missing but can still be detected on the edge of another nearby beam). Lastly, there are 4 MSPs, of which 3 have poor sensitivity and 1 has a period $< 3$~ms, which was a range that was ignored in the candidate sifting process.

Despite further investigation of potential common properties among those pulsars (e.g., fast rotation or high DM), we did not find any; these pulsars span a wide range of parameter space, as shown in Table \ref{tbl-missed-pulsar-parameters}. Hence, their non-detections are most likely due to our sifting process, which removed candidates that appear in fewer than 10 DM trials. This filters out pulsars with a detection S/N that is too faint to be detected in 10 DM trials. The same is true for MSPs, whose signal can easily become smeared if they have narrow duty cycles, even if they are largely above the nominal detection threshold.

Despite the missing pulsars, the overall larger number of pulsar detections in this reprocessing (\nviewerdetected) compared to the original survey (\norigdetected) generally reinforces the idea that motivated this reprocessing, which is that multiple searches of pulsar data sets are worthwhile, since one search package (with a particular set of search parameters and capabilities) may miss a pulsar while another may detect it (e.g., \citealt{kel+09, ekl+13, kek+13, kls+15, mbc+19, sbb+23, sbb+25}). 

\subsection{A New LMC Pulsar, PSR \newlmcpulsar} \label{sec:new-lmc-pulsar}

\begin{figure}
    \centering
    \includegraphics[width=\linewidth]{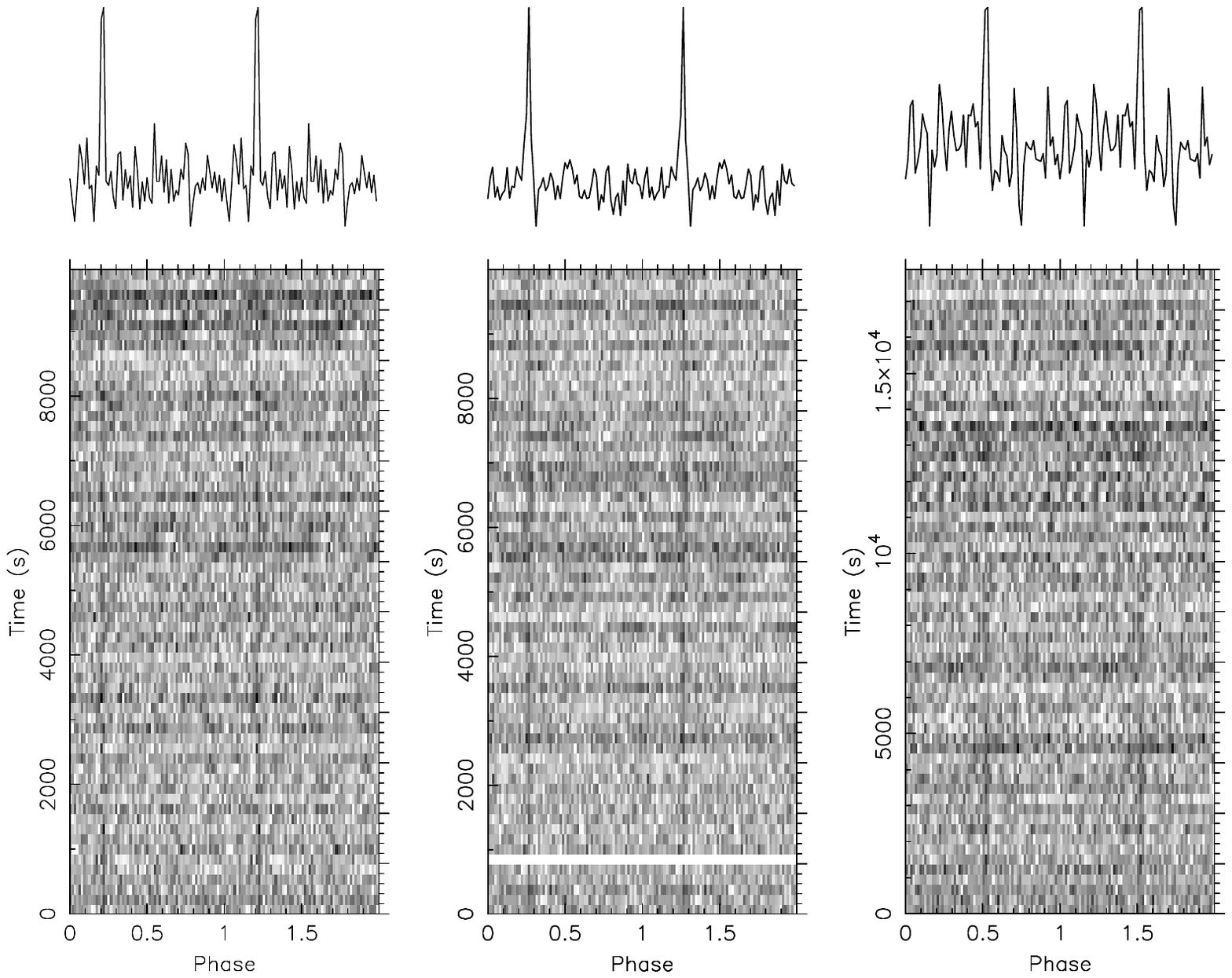}
    \caption{Folded profile detections of three detections of the new LMC pulsar PSR \newlmcpulsar, which was discovered in observations targeting PSR B0540$-$69 (see Table \ref{tbl-3}). Each subplot shows pulse phase on the horizontal axis (repeated twice for visual clarity) and time sub-integrations on the vertical axis. The summed profile is shown at the top of each subplot. The two rightmost plots show some possible variability over the course of the integration, which produces a range of flux density estimates shown in Table \ref{tbl-3} that were obtained from the folded S/N values. 
    }
    \label{fig:lmc-psr-detects}
\end{figure}

During the reprocessing, we discovered one new pulsar in three long targeted observations of the X-ray pulsar PSR B0540$-$69 that were part of the survey archive. This pulsar, PSR \newlmcpulsar\, has a spin period of \lmcpulsarperiod\ and a DM of \lmcpulsardm\ (see Table \ref{tbl-3}). This DM indicates it is located in the LMC, given the range of DMs observed for LMC pulsars and the expected maximum Galactic DM along the line of sight of 50-60~pc~cm$^{-3}$ \citep{cl02, ymw17}. Folded profile detections are shown in Fig. \ref{fig:lmc-psr-detects}. Its DM and period indicate that PSR \newlmcpulsar\ is not associated with PSR B0540$-$69, the original target of each of these observations, and that it is a new discovery.

\begin{figure}
    \centering
    \includegraphics[width=1\linewidth]{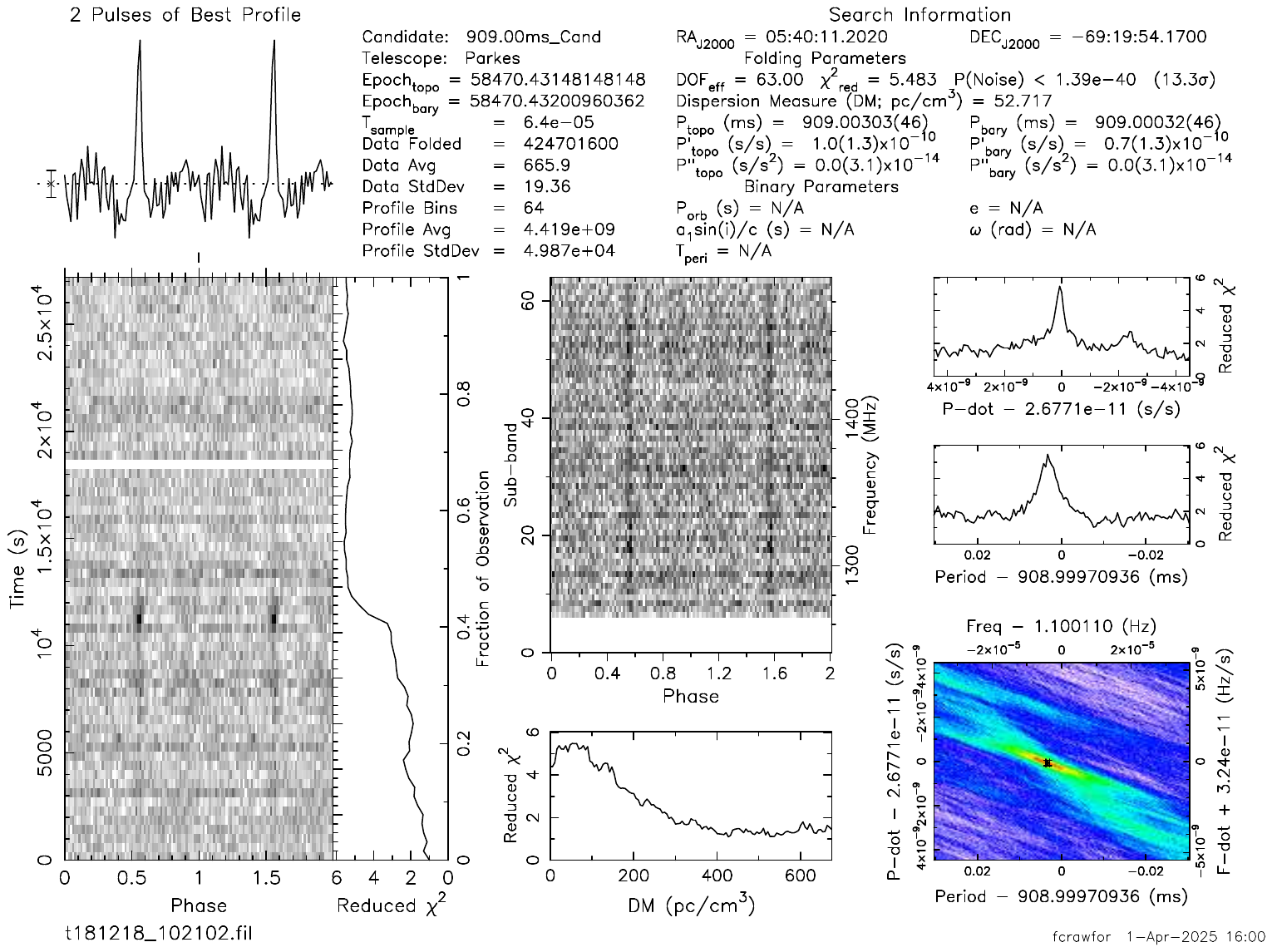}
    \includegraphics[width=1\linewidth]{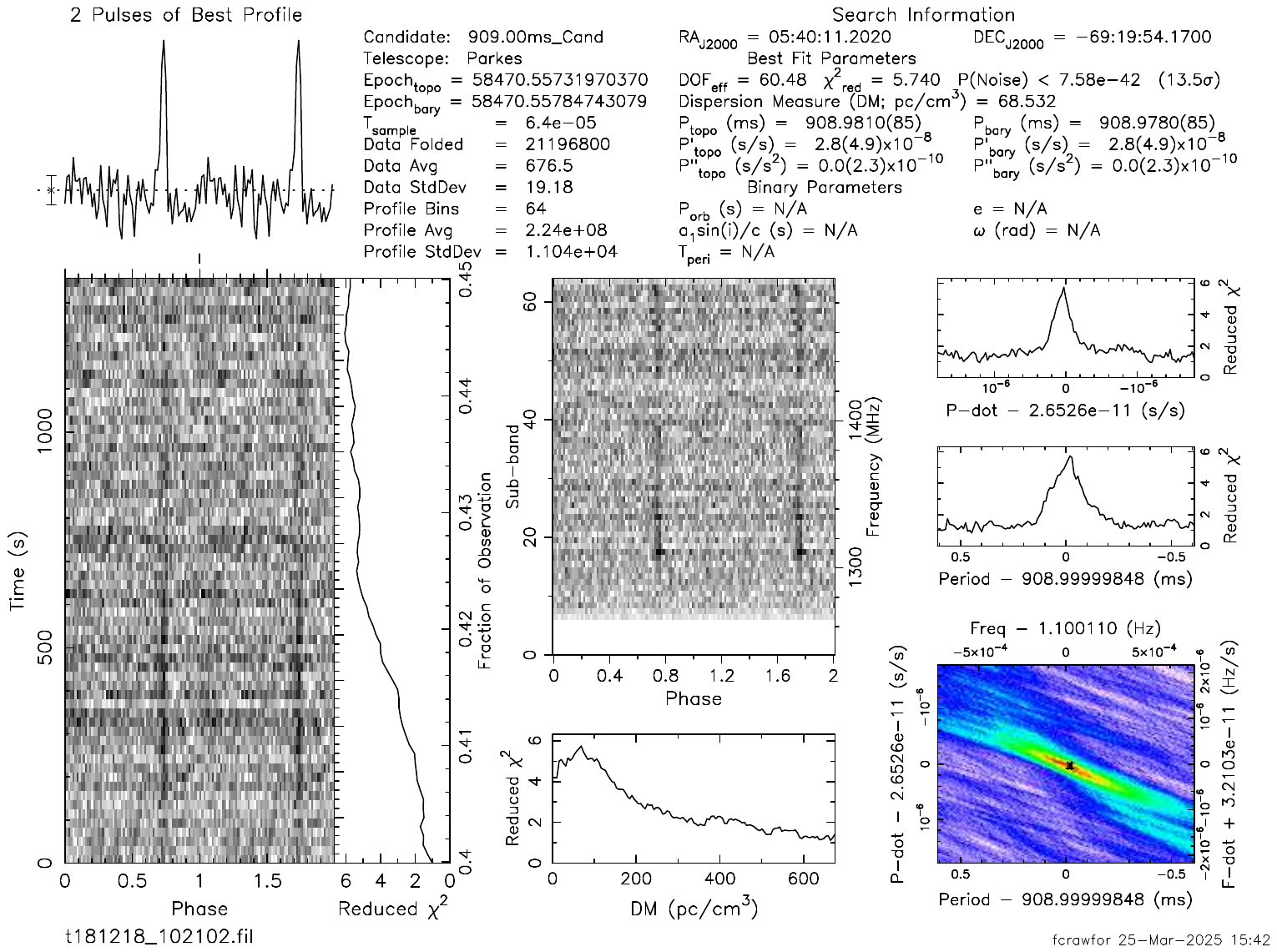}    
    \caption{
    (Top) Detection of \newlmcpulsar\ in a 1400~MHz 7.6 hr Parkes observation obtained from the CSIRO archive. This was the only other detection (apart from the three PKS70 detections) out of 47 archived observations that overlapped the beam area of the original PKS70 detections. The pulsar exhibits high variability on a timescale of tens of minutes, and therefore, no accurate flux density estimate is possible from this detection. (Bottom) A fold of the 5\% of the integration where the pulsar was brightest. The signal-to-noise is about the same as in the top fold. There is an indication of variability across the band in this sub-integration on a 100~MHz scale, unlike the full fold, in which no variability is indicated. The best-fit DM is somewhat lower than the DM found on the 436~MHz detections, but it is not well-localized in this observation (see Fig. \ref{fig:lmc-psr-dm-trials}).
    }
\label{fig:lmc-psr-prepfolds}
\end{figure}

\begin{figure}
    \centering
    \includegraphics[width=0.8\linewidth]{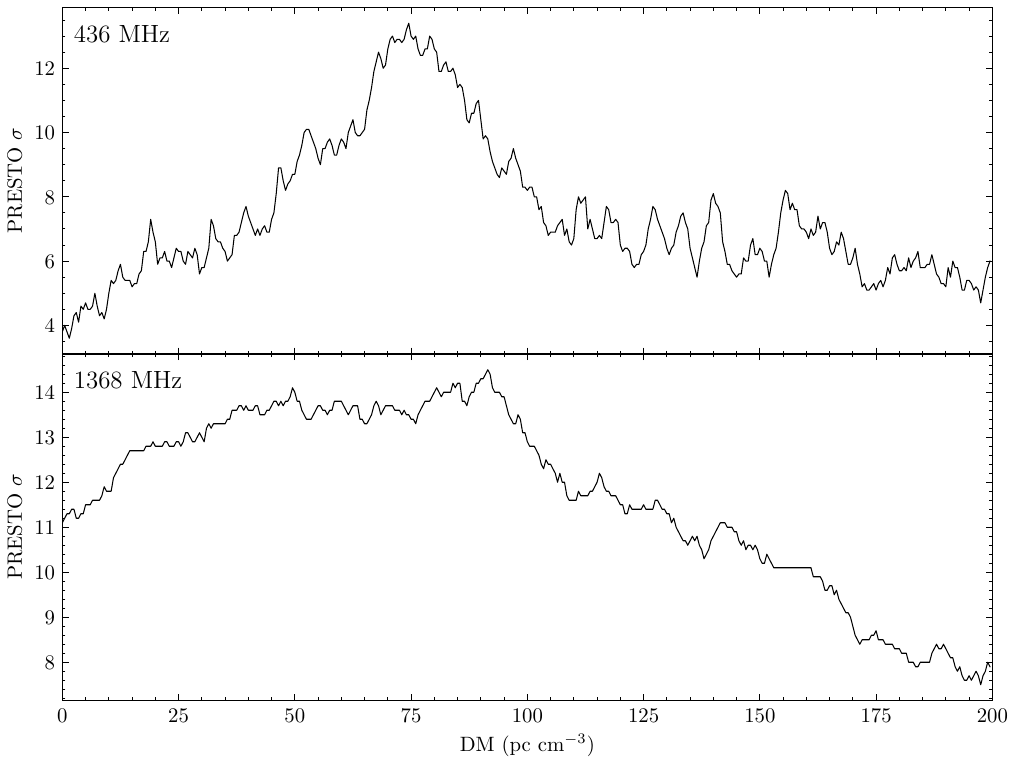}
    \caption{
    Signal strength (from the sigma value in prepfold) vs. trial DM for folds of PSR J0540$-$69 at the discovery periods. The top and bottom plots show folds at a range of DM trials for both the 436~MHz and 1380~MHz observations, respectively. The DM is more well-localized at 436~MHz (but the DM cannot be constrained better than about 5 or 10~pc~cm$^{-3}$ or so), while the DM at the 1368~MHz DM is not well localized. This is accounted for by the higher frequency and less of a smearing effect when folding at an offset DM (especially since the intrinsic pulse width in the fold is relatively wide given its 909~ms period.
    }
\label{fig:lmc-psr-dm-trials}
\end{figure}

A search of two additional PKS70 beams from the archive that covered the position did not show the pulsar. One of these non-detections was from a 10 hr integration at 600 MHz. We subsequently downloaded every other beam available in the CSIRO archive that pointed within 20 arcmin of PSR 0540$-$69; this is the half power beamwidth of the PKS70 survey beams, and our detections constrained the new pulsar position to within this radius. Each of these beams was dedispersed and folded at DMs and periods near the discovery values. Of the 45 observations searched (excluding the two PKS70 non-detections mentioned above), only one revealed a detection. This detection was in a 1368~MHz beam that targeted PSR B0540$-$69 for 7.6 hr and is shown in Fig. \ref{fig:lmc-psr-prepfolds}. This further constrains the position of PSR \newlmcpulsar\ to within 7 arcmin of PSR B0540$-$69 given this is the Parkes beam radius at this frequency. 

As seen in Fig. \ref{fig:lmc-psr-prepfolds}, the pulsar's brightness varies greatly in time over this 7.6 hr integration on a time-scale of tens of minutes. The central part of the observation is where the pulsar is brightest: folding just the brightest 5\% of the data (1360 s, or 23 min) shows a detection almost as strong as the fold of the full integration. Note that the variability is not noticeable across the 256~MHz bandwidth in the full integration. Folding just the brightest 5\% of the integration does show some possible flux variability across the bandwidth, with a scale of order 100 MHz, which is also shown in the frequency panel of Fig. \ref{fig:lmc-psr-prepfolds}. The fact that the pulsar appeared in some L-band beams but not in most of them illustrates the highly variable nature of this pulsar. This is reinforced by the variability seen in the single 1368~MHz detection.

PSR \newlmcpulsar\ has also not been detected in the recent high-resolution TRAMPUM LMC survey \citep{plg+24}. By retrieving the beam positions from the MeerKAT SARAO archive\footnote{\url{https://archive.sarao.ac.za/}} and matching them with the detection position of PSR \newlmcpulsar, it can be seen that the pulsar was covered by the TRAMPUM survey region, but at the edge of the nearest beam. Owing to this, and its high variability, PSR \newlmcpulsar\ may not be present in the MeerKAT observations. 

Integrated pulse profiles at both 436~MHz and~1368~MHz show a single pulse component of moderate width (6\% of the pulse period), suggesting that there is no profile evolution between these frequencies. 

We used the 1368~MHz detection to estimate a rough period derivative for the pulsar by comparing the period of the first 436~MHz detection with the (much later) 1368~MHz detection. This period derivative was about $3 \times 10^{-15}$, with a nominal uncertainty of about 30\% in this value, as obtained from the uncertainties in the fitted periods. This value is typical for long-period pulsars, suggesting that this pulsar is not remarkable compared to the observed pulsar population.  

The variability in the 1368~MHz detection (and the 22 other non-detections in observations taken at or near this frequency) precludes an accurate estimate of the flux density at 1400~MHz. Thus, we cannot reliably estimate a radio spectral index for PSR J0540$-$69 from these observations. This is not the case at 436 MHz, where the detections are less variable on the time scale of the integrations. We can therefore estimate the flux density at this frequency.

Given the detection at 1368~MHz and its constraint on the position of the pulsar, we can reasonably assume that the pulsar is located near the beam center (within 7 arcmin) relative to the PKS70 20 arcmin beam radius. Thus, we would not expect significant attenuation in the PKS70 detections from a beam offset position.

Using the quoted flux density limit at high Galactic latitudes of 3~mJy from the original survey \citep{mld+96} for a S/N threshold of 8, we scale the sensitivity to account for several factors: the different integration times, the different pulsed duty cycle (6\% measured for PSR \newlmcpulsar\ vs. 8\% assumed for the survey average), the S/N of the detections, and the different sky temperature at the observed location of PSR \newlmcpulsar. For this last point, the original paper \citep{mld+96} indicated a receiver temperature of 50~K and a sky temperature of 25~K (away from the Gal plane), giving a system temperature of 75~K. 

The \citet{hss+82} sky temperature map gives a sky temperature of 37~K at 436~MHz at the position of PSR \newlmcpulsar, which gives a system temperature of 87~K. This ratio of system temperatures represents a decrease in sensitivity by a factor of 1.16 for the \newlmcpulsar\ observations relative to that assumed for the survey. However, the measured, narrow pulsed duty cycle of 6\%, compared to the 8\% assumed in the survey sensitivity calculation, yields an increase in sensitivity of 1.17, which almost exactly cancels out the effect of the larger system temperature. This leaves the scaling of S/N and integration time factors as the two main contributing factors in determining the flux density.

From this scaling, we obtained flux density values of between 0.4 and 0.7~mJy (see Table \ref{tbl-3}). We use 0.6~mJy as a reasonable estimate based on these values.

\vspace{1cm}
\subsection{Comparison with Other LMC Radio Pulsars}

\begin{figure}
    \centering
    \includegraphics[width=1.0\linewidth]{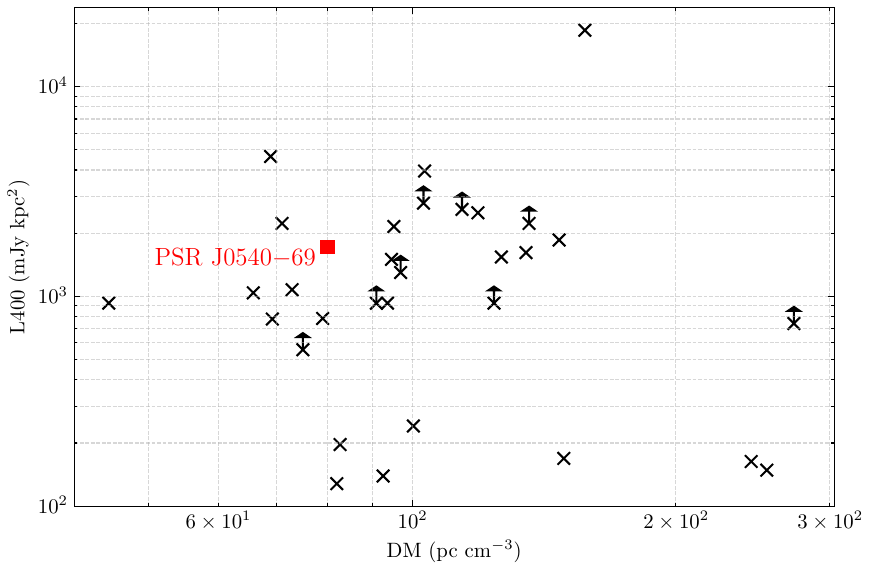}
    \caption{Estimated 400~MHz luminosity vs.~DM for the 31 currently known radio pulsars in the LMC (black crosses) and the new pulsar PSR \newlmcpulsar\ reported here (red square). This includes the 24 pulsars currently listed in the ATNF pulsar catalog (but not PSR J0537$-$6910 since it is not observed to be a radio emitter) as well as seven new discoveries reported in the LMC by \citet{plg+24} in the lower part of the plot. Two pulsars (PSRs B0456$-$69 and B0502$-$66) have cataloged 400~MHz radio luminosities, while the remainder have 1400~MHz values that were scaled to 400~MHz, assuming a spectral index of $-1.6$ \citep{jvk+18}. The 1400~MHz flux densities (or lower limits, indicated by arrows in several cases) were taken from the ATNF pulsar catalog \citep{mht+05} and from Tables 1 and 4 of \citet{plg+24}. 
    }
    \label{fig:lmc-lum-dm}
\end{figure}

Fig. \ref{fig:lmc-lum-dm} shows the estimated 400~MHz luminosity vs. DM for the known radio pulsar population in the LMC as well as the PSR \newlmcpulsar. To estimate the 400~MHz luminosity for PSR \newlmcpulsar, we used the estimated flux density of 0.6~mJy at 436~MHz (see above) and scaled it to 400~MHz using an assumed spectral index of $-1.6$, which is representative of the known pulsar population \citep{jvk+18}, (giving 0.69~mJy, an almost 15\% increase). For a 50 kpc distance from the LMC \citep{pgg+13}, this gives a 400~MHz luminosity of 1725~mJy~kpc$^{2}$. The 400~MHz luminosity values for the rest of the known LMC pulsars were obtained in all but two cases\footnote{PSRs B0456$-$69 and B0502$-$66 already have cataloged 400~MHz luminosities \citep{mht+05}.} by scaling the cataloged 1400~MHz values using an assumed spectral index of $-1.6$ (see above). The ATNF catalog records 24 LMC pulsars (excluding PSR J0537$-$6990, which is not a radio emitter), while seven new LMC pulsar discoveries were recently reported by \citet{plg+24}, resulting in 31 pulsars, plus the radio pulsar we report here. According to Fig.~\ref{fig:lmc-lum-dm}, \newlmcpulsar\ has an estimated DM and radio luminosity that is in the range of other LMC pulsars. 

\section{Conclusions} \label{sec:con}

We have presented the results of the reprocessing of the archived PKS70 survey using PRESTO. We searched an extended DM range and performed an acceleration search on all of the archival data. We detected \nbluegreen\ known pulsars, of which \ngreen\ were detected in the original survey and \nblue\ were not but could in principle have been detected at that time. Among these \nblue\ known pulsars, we also detected several highly accelerated pulsars that may have been missed in the original survey due to the absence of an acceleration search. We missed \nred\ (of the \norigdetected) pulsars that were originally detected in the survey, due in part to portions of the survey data that are missing from the survey archive \hl{and our early stage candidate sifting method}. 

We have also discovered a new LMC pulsar, PSR \newlmcpulsar, in three archival PKS70 436~MHz survey beams that targeted PSR B0540$-$69 in the LMC. PSR \newlmcpulsar\ has a spin period of \lmcpulsarperiod\ and a DM of \lmcpulsardm. The pulsar's DM suggests that it resides in the LMC. In addition to PKS70 observations, one further detection of the pulsar was made at 1368~MHz in an archival observation of PSR B0540$-$69. This detection revealed significant flux variability on a timescale of tens of minutes. This may account for why it was not seen in other archival beams targeting PSR B0540$-$69. We do not have timing information to allow for further investigation of its properties and localization beyond the 1368~MHz beam size, but the difference in spin periods at the different epochs suggests that its spin-down rate is not unusual. Planned future work with this data set includes conducting a search for long-period pulsars using the FFA algorithm \citep[e.g.,][]{mbs+20}.

\begin{acknowledgments}
We thank the students at the following institutions for their assistance with classifying pulsar candidates using the candidate viewer: Franklin and Marshall College, Kumamoto University, Hillsdale College, Kenyon College, the University of Puerto Rico at Mayagüez, Reed College, Vanderbilt University, Oregon State University, and the University of Wisconsin-Milwaukee. \hl{We also thank the anonymous referee for insightful and helpful comments that helped improve the final manuscript. }Murriyang, CSIRO’s Parkes radio telescope, is part of the Australia Telescope National Facility (\url{https://ror.org/05qajvd42}), which is funded by the Australian Government for operation as a National Facility managed by CSIRO. This paper includes archived data obtained through the Parkes Pulsar Data archive on the CSIRO Data Access Portal (\url{http://data.csiro.au}). This work was supported in part by the National Science Foundation (NSF) Physics Frontiers Center award Nos. 1430284 and 2020265, and used the Franklin and Marshall College compute cluster, which was funded through NSF grant 1925192. 
\end{acknowledgments}

\bibliography{refs}{}
\bibliographystyle{aasjournal}

\begin{deluxetable}{rcccc}
\tablecaption{Dedispersion Plan for Survey Search.}
\tablewidth{0pt}
\tablehead{
\colhead{DM range} \vspace{-0.1cm} &  \colhead{DM step} & \colhead{Number of DM trials} & \colhead{Downsampling}  
}
\startdata
   0.0 - 85.0  &  0.2  & 425 &  1 \\
  85.0 - 132.1 &  0.3  & 157 &  2 \\
 132.1 - 198.6 &  0.5  & 133 &  3 \\
 198.6 - 354.6 &  1.0  & 156 &  5 \\ 
 354.6 - 463.6 &  1.0  & 109 &  8 \\
 463.6 - 773.6  &  2.0  & 155 & 12 \\ 
 773.6 - 1723.6 &  5.0  & 190 & 24 \\ 
1723.6 - 3003.6 & 10.0  & 128 & 48 \\
\enddata 
\tablecomments{Parameters obtained from the DDplan.py script in PRESTO. All DM values are in pc~cm$^{-3}$.} 
\label{tbl-ddplan}
\end{deluxetable}

\begin{deluxetable}{lc}
\tablecaption{PKS70 Survey Parameters}
\tablewidth{0pt}
\tablehead{
\colhead{Parameter} \vspace{-0.1cm} &  \colhead{Value}  
}
\startdata
Central Observing Frequency (MHz) & 436 \\
Bandwidth (MHz)                   & 32 \\
Number of Channels                & 256 \\
Sampling Time ($\mu$s)            & 300  \\
Integration Time Per Pointing (s) & 157.3 \\
Gain                              & $0.64~{\rm K\ Jy^{-1}}$ \\
System Temperature          & $\sim $50 $_{\rm (receiver)}$ + 25 $_{\rm (sky)}$ \\
\enddata 
\tablecomments{See \citet{mld+96}.} 
\label{tbl-1b}
\end{deluxetable}

\begin{deluxetable}{lccc}
\tablecaption{Pulsar Detection Numbers}
\tablewidth{0pt}
\tablehead{
\colhead{Description} \vspace{-0.1cm} & \colhead{Total} & \colhead{Gridded} &  \colhead{Targeted}
}
\startdata
Pulsars detected in original survey \citep{mld+96, lml+98} & \norigdetected\ & $-$ & $-$ \\
Pulsars detected in this reprocessing & \nviewerdetected\ & 256 & 103  \\ 
Pulsars detected both in reprocessing and by the original survey & \nvieweroriginaldetected\ & 219 & 46 \\  
Pulsars detected by the original survey only but not in this reprocessing & 
\noriginaldetectedonly\ & $-$ & $-$ \\
Pulsars detected in this reprocessing only but not by the original survey &
\nblue\ & 37 & 57 \\
\enddata 
\tablecomments{Number of pulsars blindly detected through our reprocessing from both gridded and targeted beams, as well as the total number (i.e., gridded + targeted). Numbers refer to known pulsars only (i.e., not including the new LMC pulsar, PSR \newlmcpulsar).} 
\label{tbl-1}
\end{deluxetable}

\begin{deluxetable}{lcccc}
\tablecaption{Pulsar Detection Statistics}
\tablewidth{0pt}
\tablehead{
\colhead{Description} \vspace{-0.1cm} &  \colhead{Number of Beams} & \colhead{Both} & \colhead{PKS70$_{\rm New}$} & \colhead{PKS70$_{\rm Orig.}$}  
}
\startdata
Gridded Survey beams        & \gridbeams  & 219     &  37 & 33 \\
Targeted (confirmation) beams  & \otherbeams &  46     &  57 &  0 \\
Total                       & \totalbeams & \ngreen & \nblue & \nred \\
\enddata 
\tablecomments{Numbers do not include the new LMC pulsar we discovered in the reprocessing, PSR \newlmcpulsar. PKS70$_{\rm New}$ and PKS70$_{\rm Orig.}$ are the number of pulsars detected only by this work and the original survey \citep{mld+96, lml+98}. The number of pulsars detected by both works is shown under ``Both".}
\label{tbl-1a}
\end{deluxetable}

\begin{turnpage}
    \begin{deluxetable}{lll}
    \tablecaption{\nred\ Pulsars Detected in the Original Processing But Not in the Reprocessing}
    \tablewidth{0pt}
    \tablehead{
    \colhead{PSR} \vspace{-0.1cm} & \colhead{Refolding Status}  &  \colhead{Notes and Reason for Non-detection} 
    }
    \startdata
    \textbf{J0034-0534} & \textbf{Detected} & \textbf{MSP} \\
    J0211-8159 & No nearby beams & Missing beams \\
    J0540-7125 & No detection & No detection after folding \\
    B0621-04 & No detection & No detection after folding \\
    B0903-42 & No nearby beams & Missing beams \\
    B0950-38 & No nearby beams & Missing beams \\
    B1016-16 & No nearby beams & Missing beams \\
    \textbf{J1024-0719} & \textbf{Detected} & \textbf{Flux density near survey threshold, MSP}    \\
    B1105-59 & No detection & No detection after folding \\
    \textbf{J1126-6942} & \textbf{Detected} & \textbf{Missing the closest pointing, but detected on another nearby beam} \\
    \textbf{J1159-7910} & \textbf{Detected} & \textbf{Original survey detection S/N near survey threshold }    \\
    B1254-10 & No nearby beams & Missing beams \\
    \textbf{B1309-12} & \textbf{Detected} & \textbf{Flux density near survey threshold, located near edge of the closest beam, original survey detection S/N near survey threshold}     \\
    \textbf{J1332-3032} & \textbf{Detected} & \textbf{Missing the closest pointing, but detected on another nearby beam} \\
    J1403-7646 & No detection & No detection after folding \\
    B1454-51 & No detection & No detection after folding \\
    \textbf{B1600-27} & \textbf{Detected} & \textbf{Located near edge of the closest beam}     \\
    B1630-59 & No nearby beams & Missing beams \\
    B1657-13 & No nearby beams & Missing beams \\
    B1706-16 & No detection & No detection after folding \\
    B1717-29 & No nearby beams & Missing beams \\
    B1732-07 & No nearby beams & Missing beams \\
    \textbf{B1737-30} & \textbf{Detected} & \textbf{Located near edge of the closest beam, original survey detection S/N near survey threshold}     \\
    \textbf{B1738-08} & \textbf{Detected} & \textbf{Missing the closest pointing, but detected on another nearby beam} \\
    B1740-13 & No nearby beams & Missing beams \\
    \textbf{J1744-1134} & \textbf{Detected} & \textbf{MSP}    \\
    B1828-60 & No nearby beams & Missing beams \\
    B1841-04 & No detection & No detection after folding \\
    B1844-04 & No nearby beams & Missing beams \\
    \textbf{J1911-1114} & \textbf{Detected} & \textbf{MSP, located near edge of the closest beam}     \\
    B1940-12 & No nearby beams & Missing beams \\
    \textbf{J1940-2403} & \textbf{Detected} & \textbf{Original survey detection S/N near survey threshold }    \\
    B2043-04 & No nearby beams & Missing beams
    \enddata 
    \tablecomments{A total of \noriginaldetectedonly\ pulsars are on this list. Pulsar parameters obtained from the ATNF pulsar catalog \citep{mht+05}.}
    \label{tbl-2}
    \end{deluxetable}
\end{turnpage}

\begin{deluxetable}{lcccccc}
\tablecaption{Parameters for 12 Pulsars Missed in a Blind Search but Detected by Refolding}
\tablewidth{0pt}
\tablehead{
\colhead{PSR} \vspace{-0.1cm} & \colhead{$P$}  &\colhead{$\dot P$}  &\colhead{DM}  &\colhead{$S_{\rm 400}$}  &  \colhead{PKS70$_{\rm orig}$ S/N} & \colhead{Reference for } \\
\colhead{} \vspace{-0.1cm} & \colhead{(ms)}  &\colhead{(s/s)}  &\colhead{(pc cm$^{-3}$)}  &\colhead{(mJy)}  &  \colhead{} & \colhead{$P$, $\dot P$, DM, and $S_{\rm 400}$}
}
\startdata
J0034$-$0534 & 1.88   & 4.97$\times 10^{-21}$  & 13.8 & 24 & 16 & 1 2, 3 \\
J1024$-$0719 & 5.16   & 1.85$\times 10^{-20}$ & 6.5 & 4.6 & 12 & 4, 5 \\
J1126$-$6942 & 579.42 & 3.29$\times 10^{-15}$  & 55.3 & 9 & 17 & 6, 7 \\
J1159$-$7910 & 525.08 & 2.81$\times 10^{-15}$ & 59.3 & 6 & 10 &  6, 9 \\
B1309$-$12   & 447.52 & 1.51$\times 10^{-16}$  & 36.2 & 4.0 & 8.4 & 9, 10, 11 \\
J1332$-$3032 & 650.43 & 5.60$\times 10^{-16}$   & 15.1 & 9 & 13 & 6, 12 \\
B1600$-$27   & 778.32 & 3.01$\times 10^{-15}$ & 46.0 & 20.0 & 24.5 & 9, 11, 13 \\
B1737$-$30   & 607.07 & 4.66$\times 10^{-13}$ & 151.9 & 24.6 & 9 & 9 11, 14 \\
B1738$-$08   & 2043.09 & 2.27$\times 10^{-15}$  & 75.3 & 29 & 16.5 & 9, 11, 13 \\
J1744$-$1134 & 4.07   & 8.93$\times 10^{-21}$ & 3.1 & 18 & 22 & 4, 5, 15 \\
J1911$-$1114 & 3.63   & 1.40$\times 10^{-20}$ & 31.0 & 15 & 13 & 3, 16, 17, 18 \\
J1940$-$2403 & 1855.23 & $-$ & 63.3 & $-$ & 9 & 6 8
\enddata 
\tablecomments{
$P$, $\dot P$, DM, and $S_{\rm400}$ were taken from the ATNF pulsar catalog \citep{mht+05} and their references are indicated in the last column. PKS70$_{\rm orig}$ S/N is the detection S/N reported in the original PKS70 survey \citep{mld+96, lml+98}. 
}
\tablerefs{(1) \cite{bhl+94}; (2) \cite{aaa+10b}; (3) \cite{stc99}; (4) \cite{bjb+97}; (5) \cite{abb+23}; (6) \cite{lml+98}; (7) \cite{lbs+20}; (8) \cite{dsb+98}; (9) \cite{kjk+24}; (10) \cite{dtws85}; (11) \cite{lylg95}; (12) \cite{hlk+04}; (13) \cite{mlt+78}; (14) \cite{cl86}; (15) \cite{tbms98}; (16) \cite{llb+96}; (17) \cite{dcl+16}; (18) \cite{sbm+22}. }
\label{tbl-missed-pulsar-parameters}
\end{deluxetable}

\begin{deluxetable}{ccccccc}
\tablecaption{Survey Archive Detections of the New LMC Pulsar PSR \newlmcpulsar}
\tablewidth{0pt}
\tablehead{
\colhead{Obs Freq.} \vspace{-0.1cm} & \colhead{Observation} \vspace{-0.1cm} &  \colhead{Integration Time} & \colhead{Spin Period} & \colhead{DM} & \colhead{S/N} & \colhead{$S_{436}$} \\
\colhead{(MHz)} & \colhead{MJD} & \colhead{(hr)} & \colhead{(ms)} & \colhead{(pc cm$^{-3}$)} & \colhead{} & \colhead{(mJy)}  
}
\startdata
436  & 48633.4 & 2.8 & 908.9976(9)  &  78.8 & 11.6 & 0.54 \\
     & 48634.4 & 2.8 & 908.9976(9)  &  76.4 & 15.3 & 0.72 \\
     & 48743.2 & 5.0 & 908.9964(4)  &  84.2 & 12.5 & 0.44 \\ 
\hline
1368 & 58470.4 & 7.6 & 909.0003(5)  &  52.7 & 13.3 & $-$ 
\enddata 
\tablecomments{S/N was taken from the reported prepfold sigma in the prepfold plot. The 436~MHz flux density was estimated from the S/N and integration time (\hl{see Section \ref{sec:new-lmc-pulsar}}). The formal uncertainty from the fold in the last digit of the period is indicated by the figure in the parentheses. In all cases, the best-fit period derivative in the detection was consistent with zero. All three beams were centered on the position of PSR B0540$-$69. Folded profile detections are shown in Fig. \ref{fig:lmc-psr-detects}.} 
\label{tbl-3}
\end{deluxetable}

\end{document}